\documentclass{aa}
\usepackage{txfonts} 
\usepackage{graphicx} 
\usepackage{amssymb}
\usepackage{natbib} 
\bibpunct{(}{)}{;}{a}{}{,}
 \let\DS=\displaystyle 
\newcommand{\nFn}{\nu\mathcal{F}_{\nu}}

\newcommand{\D}{{\rm d}}

 \newcommand{\mscr}[1]{\bgroup\mathcal
#1\egroup} \newcommand{\tmop}[1]{\bgroup\rm #1\egroup}
\begin{document}
%% % Title %
\title{%%
	{\Large\mdseries Letter to Editor}\medskip\\ 
    On the dependence of the spectral parameters on the observational
    conditions in homogeneous time dependent models of the TeV blazars
    }
\titlerunning{Time dependent homogeneous TeV blazars SED}
%%
%% Author and affiliation information.
%%
\author{Ludovic Saug\'e\inst{1,2} \and Gilles Henri\inst{2}}
\institute{
    Institut de Physique Nucl\'eaire de Lyon -- UCBL/IN2P3-CNRS -- 
    4 rue Enrico-Fermi, F-69622, Villeurbanne cedex, France
    \and
    Laboratoire d'Astrophysique de Grenoble -- Universit\'e Joseph-Fourier --
    BP~53, F-38041 Grenoble, France 
    }
\offprints{Ludovic Saug\'e \email{ludovic.sauge@obs.ujf-grenoble.fr}} %
\date{Received --- ; accepted --- }%%
\abstract
{}{   
    Most of current models of TeV blazars emission assume a Synchrotron
    Self-Compton mechanism where relativistic particles emit both synchrotron
    radiation and Inverse Compton photons. For sake of simplicity, these
    models usually consider only steady state emission. The spectral features
    are thus only related to the shape of the particle distribution, and do
    not depend on the timing of observations.
}{   
    In this letter, we study the effect of, firstly, the lag between the
    beginning of the injection of the fresh particles and the trigger of the
    observation, and secondly, of a finite injection duration. We illustrate
    these effects considering an analytical time-dependent model of the
    synchrotron emission by a monoenergetic distribution of leptons.
}{   
    We point out that the spectral shape can  be in fact very dependent on
    observational conditions if the particle injection term is
    time-dependent, particularly taking into account the effect of the time
    averaging procedure on the final shape of the SED. Consequences on the
    acceleration process are also discussed.
}{}    
 %% ----------------------------------------------------------------------
\maketitle
%%
%% ----------------------------------------------------------------------
    \section{Introduction}
%% ----------------------------------------------------------------------
%%
TeV blazars  belong to the radio-loud active galactic nuclei (AGN) and are
characterized by non-thermal continuum spectrum, optical polarization, flat
radio spectrum and strong variability in all frequency bands. The broad band
of the SED which extends from radio up to the TeV range, consists typically in
two bumps. The low energy one --- peaking in the $0.1-100\rm\ keV$ domain ---
is commonly attributed to synchrotron emission of ultra-relativistic leptons.
In the \emph{Synchrotron-Self-Compton process} (SSC) the second component is
attributed to the inverse Compton (IC) scattering of synchrotron photon field
by the same population of leptons.  Emission of TeV photons provides evidence
for the presence of particles at roughly the same energy if we suppose that
they radiate in the IC Klein-Nishina regime \citep{bg70}.  Therefore, it
requires an extremely efficient particle acceleration mechanism at work in the
close environment of supermassive black hole.

Extended observations of TeV blazars like Mrk\ 501 ($z_s=0.034$,
\citealp{quinn96}) show that synchrotron component can be very variable in the
UV/X-ray range. The simplest approach is the well-known homogeneous or
one-zone modeling. It considers the emission of a relativistic moving blob of
plasma filled by a tangled magnetic field, where ultra-relativistic particle
are injected and can cool freely (via both synchrotron and inverse Compton
scattering process). Generally, in these models, the acceleration mechanism is
not consistently taken into account and the resulting \emph{energy
distribution function} (EDF) of injected particles is prescribed. If we
supposed that the acceleration zone could be dissociated from the radiative
one, then we must distinguish two different kinds of EDF : the \emph{injected}
one and the \emph{cooled} one. In the usual one-zone model, the first is
directly the result of the acceleration mechanism. The second ensues from the
radiative cooling of particles (and possible re-acceleration mechanism). It is
mathematically speaking the solution of the standard kinetic equation which
the injected EDF is the main source term. 
The exact shape of the SED could be only understood in the framework of
time dependent modeling, depending on the detail of the cooling and injection
processes \citep{krm98,cg99}. 
The majority of SSC models are steady-state: therefore the instantaneous SED
is equal to the observed one. In this case, the authors use broken power-law
EDF as the solution of the kinetic equation (characterized by the indices
$\alpha_{\ell}$ and $\alpha_u$). Then the subsequent synchrotron spectra have
also a broken power-law shape\footnote{%% 
        Only true in the case where particle energy dynamical range,
        \textit{i.e.} the ratio $\gamma_{\rm max}/\gamma_{\rm min}$ is
        much larger than unity.  }
$\mathcal{F}_{\nu}\propto\nu^{-\beta_{\ell,u}}$ with indices\footnote{%
    At low frequency, this latter can not be steeper than
    the one produced by a monoenergetic EDF, given the lower limit
    $1/3$.
}
$\beta_{\ell,u}=(\alpha_{\ell,u}-1)/2 \leqslant 1/3$ over suitable frequency
ranges \citep{bg70}. Spectral index variations can arise from the variation of
$\alpha_{\ell,u}$ \textit{i.e.} by varying the physical conditions of the
emission zone (see \textit{e.g.} \citealp{kt04}). 

We have recently investigated another solution, still in the framework of
homogeneous modeling, but with a time dependent particle injection term
\citep[][\null, hereafter paper I]{sauge04a}.  The kinetic equation is solved
numerically taking into account \emph{(i)} the possible in-situ pair
reprocessing and  \emph{(ii)} both synchrotron and IC cooling term. The source
term of the equation is chosen as a quasi-monenergetic distribution of
particles (or ``\emph{pileup}''), injected during a finite time.
Physically, the formation of such a distribution results from the combination
of a stochastic heating via second order Fermi process and radiative cooling
\citep{hp91,s85}. Then the resulting cooled EDF is also partially a power-law,
but with a constant index value $\alpha=2$ over a dynamical range depending on
the details of the cooling process.  In this case, the instantaneous
synchrotron spectrum is a power-law of index $\beta=1/2$ and no spectral index
variation could be expected at this stage.  However, such a variation can be
obtained by averaging out the instantaneous SED over the time and considering
that the observation starting time does not necessarily coincide with the
beginning of injection phase.

In this letter we study in more detail this effect. First of all, we derive a
basic model in Sect. 2 in the idealized case of monoenergetic injected EDF
subject only to synchrotron cooling. Then we derive the global shape of the
time average synchrotron part of the SED as a function of the observational
parameters. Finally, we illustrate our approach on {Beppo}SAX 
archival data sets of the object Mrk~501 in Sect. 3, before concluding.  
%%
%% ----------------------------------------------------------------------
    \section{The model}
%% ----------------------------------------------------------------------
%%
In the following, all times are expressed in the plasma (blob) rest frame.  We
neglect the photon flight time across the source by considering evolution only
less than $R/c$ where $R$ is the transverse size of the source (see Paper I).
This issue was previously studied in detail by \citet{cg99} in the case of the
instantaneous emission.
%%
%% ----------------------------------------------------------------------
    \subsection{Energy Distribution Function of Particles}
%% ----------------------------------------------------------------------
%%
For analytical purpose, we consider the simple case of the injection of a
pure monoenergetic EDF of particle of energy $\gamma_{\max,0}$ during a
finite time $t_{\rm inj}$ 
\begin{equation}
\mathcal{Q}_{\rm inj}(\gamma;t)
=\mathcal{Q}_0\,\delta(\gamma-\gamma_{\max,0})\,\Pi(t;0,t_{\rm inj}),
\end{equation}
and we consider that the main channel of particle cooling is the synchrotron
process.  We also neglect the in-situ pair creation process which can strongly
affect the resulting cooled EDF in case of large $\gamma\gamma$-opacity (see
Paper I). We introduced the gate function $\Pi(x;x_{\rm min},x_{\rm
max})=\Theta(x-x_{\rm min})\Theta(x_{\rm max}-x)$ where $\Theta$ is the usual
Heaviside unit step function. We define the cooling time of a particle with a
Lorentz factor $\gamma$ by the usual relation $t_{\rm
cool}(\gamma)={1}/{k_{\rm syn}\gamma}$, where $k_{\rm syn}={\sigma_{\rm
Th}}/{6\pi m_ec}B^2$.  It also represents the time spent by an initial
infinite energy particle to cool down to $\gamma$. In the following, another
characteristic time value related to the previous one will be $t_{\rm
crit}=t_{\rm cool}(\gamma_{\rm max,0})$. Under these assumptions the solution
of the standard kinetic equation is analytic
: $n(\gamma;t)=n_0\gamma^{-2}\Pi(\gamma;\gamma_{\min}(t),\gamma_{\max}(t))$
where the power-law index $2$ is typical of the synchrotron radiative cooling.
Values of time-dependent bounds write
\begin{equation}
  \gamma_{\min} ( t ) = \frac{\gamma_{\max, 0}}{1 + t / t_{\tmop{crit}}} 
  \ ;\ 
  \gamma_{\max} ( t ) = \frac{\gamma_{\max, 0}}{1 + \max ( 0,
  t - t_{\tmop{inj}} ) / t_{\tmop{crit}}} .
\end{equation}
We also define the dynamical range of distribution as \emph{ie} the ratio
$r_{\tmop{dyn}}(t)= \gamma_{\max}(t)/\gamma_{\min}(t)$.
%%
%% ----------------------------------------------------------------------
    \subsection{Instantaneous synchrotron spectrum}
%% ----------------------------------------------------------------------
%%
As said before, in the case where the previous dynamical range is sufficiently
wide, the synchrotron spectrum is itself a power-law of index $1/2$ in $\nFn$.
We recall that a particle of energy $\gamma$ emits preferentially synchrotron
radiation roughly at the maximum frequency 
%% $\nu_{\tmop{smax}}(\gamma)=0.29\times{3\gamma^3\omega_{s}(\gamma)}/{4\pi}$
$\nu_{\tmop{s}}(\gamma)=0.29\times{3\gamma^3\omega_{s}(\gamma)}/{4\pi}$ 
where the relativistic synchrotron gyro-frequency writes
$\omega_{s}(\gamma)={qB}/{\gamma m_ec}$ \citep{bg70}. Then we consider a
$\delta$-approximation of the synchrotron kernel function in the optically
thin emission limit, centred around $\nu_{\tmop{s}}$.  In this case, the
synthetic synchrotron component SED can be approximated by $\nFn ( t ) = S_0
\nu^{1/2} \Pi(\nu;\nu_{\tmop{smin}}(t),\nu_{\tmop{smax}}(t))$ over the
frequency range given by
\begin{equation}\label{eq:nurange}
\left\{\begin{array}{rcl}
  \DS
  \nu_{\tmop{smax}} ( t ) &=& \DS\frac{\nu_{\tmop{smax}, 0}}{\left[ 1 + \max ( 0, t
  - t_{\tmop{inj}} ) / t_{\tmop{crit}} \right]^2},\\
  \DS
  \nu_{\tmop{smin}} ( t ) &=& \DS\frac{\nu_{\tmop{smax}, 0}}{( 1 + t /
  t_{\tmop{crit}} )^2},
\end{array}\right.
\end{equation}
where $\nu_{\tmop{smax}, 0}=\nu_{\tmop{smax}}(t=0)=\nu_{\tmop{s}}(\gamma_{\max, 0})$.
%%
%%
%% ----------------------------------------------------------------------
    \subsection{Time averaged synchrotron spectrum}
%% ----------------------------------------------------------------------
%%
%%
As previously said, the observed spectrum results from the time average of
instantaneous spectra from $t_{\rm obs}$ to $t_{\rm obs}+\Delta t_{\rm
obs}$. The time origin $t=0$ is related to the beginning of the injection
of fresh particles. One writes,
\begin{eqnarray}\label{eq:nfnobs}
 \langle \nFn \rangle_t 
  &=& \frac{S_0 \nu^{1 / 2}}{\Delta t_{\tmop{obs}}}
  \int_{t_{\tmop{obs}}}^{t_{\tmop{obs}} + \Delta t_{\tmop{obs}}} \D t\ 
  \Pi(\nu;\nu_{\tmop{smin}}(t),\nu_{\tmop{smax}}(t)),\\
  &=& \frac{S_0\nu^{1/2}}{\Delta t_{\tmop{obs}}}\, (\mathcal{I}_1+\mathcal{I}_2),
\end{eqnarray}
where the quantities $\mathcal{I}_1$ and $\mathcal{I}_2$ arise from the
splitting of the previous integral over $[t_{\rm obs},t_{\rm inj}]$ and
$[t_{\rm inj},t_{\rm obs}+\Delta t_{\rm obs}]$. After straightforward
algebra, these values depend on the position of $t_{\tmop{obs}}$ compared
to $t_{\tmop{inj}}$, and we distinguish two cases,
\begin{eqnarray*}
\bullet\ \mbox{\bfseries case A}&:&t_{\tmop{obs}} \leqslant t_{\tmop{inj}}\\
    \mathcal{I_{}}_1 & = & \max [ 0, t_{\tmop{inj}} - \max ( t_{\tmop{obs}},
    t_{\tmop{crit}} ( \nu ) ) ],\\
    \mathcal{I_{}}_2 & = & \min [ t_{\tmop{obs}} + \Delta t_{\tmop{obs}},
    t_{\tmop{inj}}+ t_{\tmop{crit}} ( \nu ) ] - \max [ t_{\tmop{inj}},
    t_{\tmop{crit}} ( \nu ) ],\\
\bullet\ \mbox{\bfseries case B}&:&t_{\tmop{obs}} \geqslant t_{\tmop{inj}}\\
    \mathcal{I_{}}_1 & = & 0,\\
    \mathcal{I_{}}_2 & = & \min [ t_{\tmop{obs}} + \Delta t_{\tmop{obs}},
    t_{\tmop{inj}}+ t_{\tmop{crit}} ( \nu ) ] - \max [ t_{\tmop{obs}},
    t_{\tmop{crit}} ( \nu ) ].
  \end{eqnarray*}
where we define $t_{\tmop{crit}}(\nu) =
t_{\tmop{crit}}\times[({\nu_{\tmop{smax},0}}/{\nu})^{1/2}-1]$ representing the
time needed by the lower bound of the synchrotron spectrum to reach the
frequency $\nu$.
%%
%% ----------------------------------------------------------------------
    \section{Results and Discussion}
%% ----------------------------------------------------------------------
    \subsection{Effects of parameters}
\begin{figure}[t]
  \resizebox{\hsize}{!}{\includegraphics[width=\hsize]{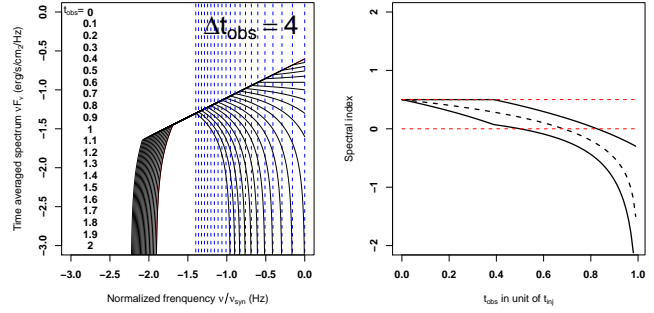}}
  \caption{%
  Influence of the value of $t_{\rm obs}$ parameter on the time averaged
  SED (\emph{left panel}) with $\Delta t_{\rm
  obs}=4 t_{\rm inj}$ and $t_{\rm crit}=0.5 t_{\rm inj}$. Vertical dashed
  lines represent location of the frequency break $\nu_{\rm break}$
  calculated for each previous curve (see eq. [\ref{eq:nubreak}]).
  Resulting local spectral indices in normalized frequency range
  $[-0.5,0]$ (upper solid curve), 
  $[-1,-0.5]$ (lower solid curve) and
  $[-1,0]$ (dashed  curve)
  are given on the right panel.}\label{fig1}
\end{figure}

%% ----------------------------------------------------------------------
%%
Using the previous relations, we show the influence of the $t_{\rm obs}$
parameter on the time averaged spectra on Fig. \ref{fig1}. On theses figures,
the time is normalized in unit of $t_{\rm inj}$ and frequency in unit of
$\nu_{\rm smax,0}$. If $t_{\rm obs}+\Delta t_{\rm obs}>t_{\rm inj}$, the
resulting SED consists of two parts (see left panel). The lower one exhibits a
power-law tail of constant index $1/2$ and results from the emission of most
energetic particle injected since the start of the observation (and not from
the beginning of the injection) and which have not yet had the time to cool.
Conversely, up to $\nu_{{\rm break}}$ the spectrum softens. In this case we
obtain from (\ref{eq:nurange}) $\nu_{{\rm break}}=\nu_{\rm smax}(t_{\rm
obs})$ or explicitly 
\begin{equation}\label{eq:nubreak}
  \nu_{\tmop{break}}(t_{\rm obs}) = \frac{\nu_{\tmop{smax}, 0}}{( 1 + t_{\tmop{obs}} /
  t_{\tmop{crit}} )^2} .
\end{equation}
In the case $t_{\rm obs}\leqslant t_{\rm inj}\leqslant t_{\rm obs}+\Delta
t_{\rm obs}$, the high energy part could be also described by a power-law with
local index $\alpha$, which implicitly depends on $t_{\rm obs}$. Combining
previous calculations, we obtain
\begin{eqnarray}
\label{eq:indexalpha}\alpha&=&
	\frac{
		\log \left[\langle\nFn\rangle(\nu_{\rm smax,0})/ \langle\nFn\rangle(\nu_{\rm break})\right]
		}{
		\log \left[\nu_{\rm smax,0}/ \nu_{\rm break}\right]
		}\\
&=& \frac{
\log \left[\left(1-t_{\rm obs}/t_{\rm inj}\right)\left(1+t_{\rm obs}/t_{\rm crit}\right)
\right]}{2\log\left(1+t_{\rm obs}/t_{\rm crit}\right)}
\leqslant
\frac 12\left(1-\frac{t_{\rm crit}}{t_{\rm inj}}\right)
\end{eqnarray}
where the upper bound is obtained for $t_{\rm obs}=0$. Note that the
spectrum is expected to be flat ($\alpha=0$) when $t_{\rm obs}=t_{\rm
inj}-t_{\rm crit}$.
%%%
\begin{figure}
  \resizebox{\hsize}{!}{\includegraphics{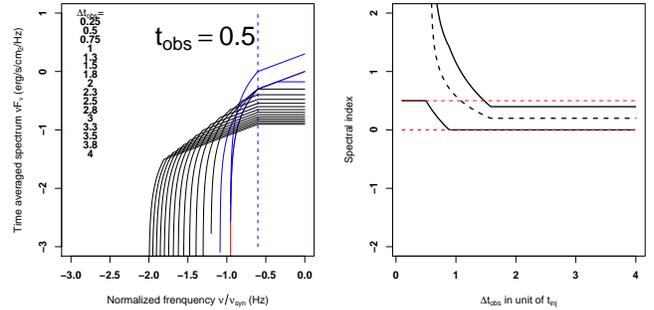}}
  \caption{%
  Same as Fig. \ref{fig1} for the influence of the $\Delta t_{\rm obs}$
  parameter with $t_{\rm obs}=t_{\rm crit}=0.5 t_{\rm
  inj}$. Considering these parameter values, we easily verify that spectral
  index above $\nu_{\rm break}=0.25\,\nu_{\rm smax,0}$ is going to be flat
  since $\Delta t_{\rm obs}\geqslant t_{\rm inj}-t_{\rm obs}=0.5$.
}\label{fig2}
\end{figure}
This mechanism gives a plausible explanation of the broken power-law or curved
shape of TeV blazars reported by many authors (see \textit{e.g.}
\citealp{fossati00a, fossati00b}). Following our study, we deduce that the low
energy synchrotron index should be always be equal to $1/2$ in $\nFn$
representation. Up to $\nu_{\rm break}$, the index value strongly depends both
on the observational characteristics and on the details of the cooling
process.  Moreover, for $t_{\rm obs}\gg t_{\rm inj}-t_{\rm crit}$, this part
of the SED strongly steepens and akin to a pure power-law. It is interesting
to note that the maximum curve naturally moves from $\nu_{\rm smax,0}$ to
$\nu_{\rm break}$ as the time increases and strongly depends on the value of
observational parameters. Right panel shows the time evolution of spectral
index calculated in three different logarithmic frequency bands, namely
$[-0.5,0]$, $[-1,-0.5]$ and $[-1,0]$.

Similarly, Fig. \ref{fig2} shows the main effects of the increase of the
$\Delta t_{\rm obs}$ parameter. In the case of $t_{\rm obs}\leqslant t_{\rm
inj}\leqslant t_{\rm obs}+\Delta t_{\rm obs}$, the spectral shape does not
evolve. As expected, only the global flux level changes (decreasing linearly
with $\Delta t_{\rm obs}$) just as the low-frequency cut-off (decreasing as
maximum system time $t_{\rm obs}+\Delta t_{\rm obs}$ increases). 
%%
%% ----------------------------------------------------------------------
    \subsection{Application to Markarian 501}
\begin{figure}[h]
  \resizebox{\hsize}{!}{\includegraphics{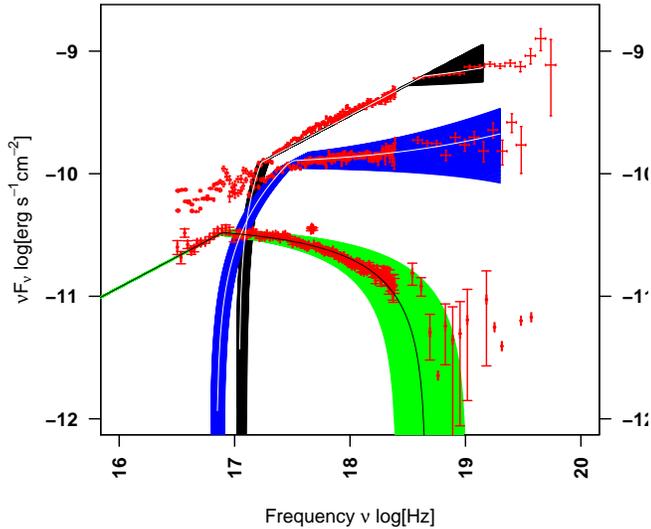}}
  \caption{%
        Fits of Markarian\ 501 UV/X-ray synchrotron spectrum for three
        different periods.
  }\label{fig3}%
\end{figure}

%% ----------------------------------------------------------------------
%%
Mrk~501 is a privileged TeV blazar target. It has been studied extensively
during many multi-wavelength campaigns and have been observed in various
spectral states.  We now test our simple approach on state extracted from
the {Beppo}SAX mission public archive\footnote{%
\texttt{http://www.asdc.asi.it/}
}
. In the following we consider three
different data sets, two from the 1997 April flaring period \citep{pian98}
and one from the 1999 June period \citep{tavecchio01}. Corresponding SED
are given in Fig. \ref{fig3}. The two first SED have been studied in Paper
I with the more complete homogeneous SSC model and using high-energy data
in order to constrain all physical quantities.

\begin{table}[h]
    \centering
    \caption{
    Parameters used in our simulations.  All times are expressed in unit of
    $t_{\rm inj}$\label{tab:param} }
    %%%%%%%%%%%%%%%%%%%%%%%%%%%%%%%%%%%%%%%%%%%%%%%%%%%%%%%%%%%%%%%%%%%%%%
    \begin{tabular}{lccccc}
    \hline\hline
            {data sample} 
            &{$t_{\rm crit}$}
            &
            &$t_{\rm obs}$
            &
            &{$\Delta t_{\rm obs}$}\\
            \cline{3-5}
            &
            &min
            &max
            &est\\
            \hline
    1997 April 7  &0.1095 & 0.68 & 0.92 & 0.8  & 0.946 \\
    1997 April 16 &0.438  & 0    & 0.5  & 0.35 & 4.2   \\
    1999 June     &0.0512 & 1.05 & 1.15 & 1.1  & 4.5   \\
    \hline
    \end{tabular}
    %%%%%%%%%%%%%%%%%%%%%%%%%%%%%%%%%%%%%%%%%%%%%%%%%%%%%%%%%%%%%%%%%%%%%%
\end{table}

All parameter values needed for the synchrotron part of the SED fits and
suitable estimations are given in Table \ref{tab:param}. Again, the times are
expressed in unit of $t_{\rm inj}$. For each period, we represent a continuous
set of curves as the function of $t_{\rm obs}$ ; also represented, our best
estimation of $t_{\rm obs}$. 
%%
%% ----------------------------------------------------------------------
    \subsubsection{1997 April period}
%% ----------------------------------------------------------------------
%%
For the 1997 April 16 (\textit{medium state}) and April 7 (\textit{high
state}) observations, we find of course values quite similar to Paper I.  Fits
are very good except for the extreme high energy part of the spectra.  The
exact form of this cut-off depends both on the exact profile synchrotron
kernel and on the exact shape of the particle cooled EDF which are rather
approximate in this work. For both periods, the beginning of the observation
takes place before the end of the injection phase and therefore the
synchrotron spectra displays a positive $\alpha$ value, as defined in
equation (\ref{eq:indexalpha}).  

First analysis done by \citet{pian98} using an one-zone model with a power-law
injection function show that the April 16 spectrum is correctly reproduced for
an \emph{injected EDF} with an index $s=1$ over a dynamical range less than
10. In the usual context of shock acceleration process, the value of the index
$s$ case, the index $s$ directly depends on the physical conditions at the
shock \citep{bo78,ps99},
\begin{equation}\label{eq:plindex}
	s=1+\frac{t_{\rm acc}}{t_{\rm esc}}=\frac{R+2}{R-1}\ >1.
\end{equation}
In the last expression, $t_{\rm acc}$ and $t_{\rm esc}$ are respectively the
acceleration and the escape characteristic time, and $R$ the shock compression
ratio. Then the value $s=1$ is highly unlikely, because it requires $t_{\rm
esc}\to\infty$, which is not consistent : particles can not escape from the
accelerator. Moreover, the narrowness of the dynamical range in this latter
case clearly pleads for a monoenergetic injection function and consequently
for a stochastic acceleration process.

In classical shock acceleration model, the difference between the 1997's
\textit{high} and \textit{medium} state is explained by difference in the
shock physical condition --- \textit{e.g.} via the compression ratio $R$ ---
leading to a different index value $s$. In the light of our scenario and as
already noted in Paper I, the difference in the shape of the spectra arises
essentially from different observational conditions. The medium state
corresponds to a previous injection observed in a later stage (with respect to
beginning of the injection phase). This is corroborated by the flatness of the
high energy part of the X-ray spectrum.
%%
%% ----------------------------------------------------------------------
    \subsubsection{1999 June period}
%% ----------------------------------------------------------------------
%%
During 1999 June period the synchrotron spectrum experienced a very steep
state.  In the context of the power-law injected EDF, this state is explained
with a large value of the index $\alpha_u$ of the upper part of the EDF,
roughly equals to $4.3$ \citep{tavecchio01}.

In the shock acceleration scenario the value of $\alpha_u$ is related to $s$
via the relation $\alpha_u=s+1$ \citep{kt04}. Then in our precise case,it
corresponds to an acceleration time quite larger than the escape one ($t_{\rm
acc}=3.3\, t_{\rm esc}$ or, in term of compression ratio $R\approx 1.9$). In
our approach, this spectral shape is obtained considering an observation time
$t_{\rm obs}$ larger than $t_{\rm inj}-t_{\rm crit}\approx t_{\rm inj}$. 
%%
%% ----------------------------------------------------------------------
    \section{Conclusion}
%% ----------------------------------------------------------------------
%%
We have exposed a simple time dependent mechanism in order to explain the
spectral shape of the synchrotron spectrum of blazar considering the injection
of a pure monoenergetic distribution of ultra-relativistic particles over a
finite time range. This latter EDF arises from stochastic acceleration
mechanism (Second order Fermi process). Spectral index and SED shape
variations can be explained by the value of the starting observation time with
respect to the beginning of the injection of fresh particles and from the
variation of the acceleration condition even if the instantaneous synchrotron
spectrum is universal with a constant spectral index equals to $1/2$. Because blazars are extremely variable sources, there are often observed in
``\emph{Target-of-Opportunity}" mode, where satellites observations are
triggered by some other instruments after some delay leading naturally to
$t_{\rm obs}\neq 0$. This
mechanism leads to complex time dependent behaviors of the spectral shape ; we
show that we can reproduce various spectral shapes from the single to broken
power-law shape, with strong or soft cut-off. This model differs from the
usual one in the context of the shock acceleration model, where the spectral
variability arises from the physical conditions at the shock. 
%%
%%
%% ----------------------------------------------------------------------
\begin{acknowledgements}
    Remarks of an anonymous referee helped to improve the content
    of this letter. LS particularly thanks Ga{\"e}lle Boudoul for her helpful
    comments and her interest. All members of the SHERPA team (Grenoble) and
    the IPNL team of the SNFactory collaboration (Lyon) are also warmly
    acknowledged. All computations and figures were performed with the free
    software \texttt{R} (\texttt{http://www.r-project.org})
\end{acknowledgements}
%% ----------------------------------------------------------------------

%% ----------------------------------------------------------------------
% THE BIBLIOGRAPHY
%% ----------------------------------------------------------------------

%% ----------------------------------------------------------------------
%%


\begin{thebibliography}{}
\bibitem[Blandford \& Ostriker(1978)]{bo78} Blandford, R.~D.~\& Ostriker, J.~P.\ 1978, \apjl, 221, L29
\bibitem[Blumenthal \& Gould(1970)]{bg70} Blumenthal G., Gould R. 1970, Rev. Mod. Phy., vol 2, 2, 237
\bibitem[Chiaberge \& Ghisellini(1999)]{cg99} Chiaberge, M., \& Ghisellini, G.\ 1999, \mnras, 306, 551 
\bibitem[Fossati et al.(2000a)]{fossati00a} Fossati, G., et al.\ 2000, \apj, 541, 153 
\bibitem[Fossati et al.(2000b)]{fossati00b} Fossati, G., et al.\ 2000, \apj, 541, 166 
\bibitem[Henri \& Pelletier(1991)]{hp91} Henri P., Pelletier  G. 1991, ApJ, 383, L7
\bibitem[Kino \& Takahara(2004)]{kt04} Kino, T. \& Takahara, F. 2004, \mnras, 349, 336
\bibitem[Kirk et al.(1998)]{krm98} Kirk, J.~G., Rieger, F.~M., \& Mastichiadis, A.\ 1998, \aap, 333, 452  
\bibitem[Pian et al.(1998)]{pian98} Pian, E.~et al.\ 1998, \apjl, 492, L17 
\bibitem[Protheroe \& Stanev(1999)]{ps99} Protheroe, R.~J.~\& Stanev, T.\ 1999, Astropart. Phy., 10, 185 
\bibitem[Punch et al.(1992)]{punch92} Punch, M., et al. 1992, Nature, 358, 477
\bibitem[Quinn et al.(1996)]{quinn96} Quinn, J., et al. 1996, \apj, 456, L83
\bibitem[Saug\'e \& Henri(2004)]{sauge04a} Saug{\' e}, L., \& Henri, G.\ 2004, \apj, 616, 136 (Paper I)
\bibitem[Schlickeiser(1985)]{s85} Schlickeiser, R.\ 1985,\aap, 143, 431
\bibitem[Tavecchio et al.(2001)]{tavecchio01} Tavecchio, F.~et al.\ 2001, \apj, 554, 725
\end{thebibliography}
\end{document}